\documentstyle[epsfig,twocolumn,aps]{revtex}
\begin{document}
\title{Statistics of the Coulomb blockade peak spacings of a silicon quantum dot}
\author{F. Simmel$^{1}$, David Abusch-Magder$^{1,2,*}$, D. A. Wharam$^{3},$\\
M. A. Kastner$^{2}$, and J. P. Kotthaus$^{1}$ \\}
\address{(1) Center for NanoScience and Sektion Physik, LMU M\"{u}nchen,\\ 
Geschwister-Scholl-Platz 1,
D-80539 M\"{u}nchen,\\
(2) Department of Physics, Massachusetts Institute of Technology, Cambridge, MA 02139,\\
(3) Institut f\"{u}r Angewandte Physik, Universit\"{a}t T\"{u}bingen,\\    
 Auf der Morgenstelle 10, D-72076 T\"{u}bingen}
\date{\today } 
\maketitle 

\begin{abstract}
We present an experimental study of the fluctuations of Coulomb
blockade peak positions of a quantum dot.  The dot is defined by
patterning the two-dimensional electron gas of a silicon MOSFET
structure using stacked gates.  This permits variation of the number
of electrons on the quantum dot without significant shape distortion.
The ratio of charging energy to single particle energy is considerably
larger than in comparable GaAs/AlGaAs quantum dots. The statistical
distribution of the conductance peak spacings in the Coulomb blockade
regime was found to be unimodal and does not follow the Wigner
surmise.  The fluctuations of the spacings are much larger than the
typical single particle level spacing and thus clearly contradict the
expectation of random matrix theory.
\end{abstract}

\noindent

The spectral properties of many quantum mechanical systems whose
classical behavior is known to be chaotic are remarkably well
described by the theory of random matrices (RMT)~\cite{mehta91}.  This
has been experimentally confirmed, for example, in measurements of
slow neutron resonances of nuclei~\cite{haq82} and in microwave
reflection spectra of billiard shaped
cavities~\cite{graef92}. Electron transport experiments performed on
semiconductor quantum dots in the Coulomb blockade (CB)
regime~\cite{cbreview} provide a further possibility to check RMT
predictions.  The classical motion of electrons in these structures
can be assumed to be chaotic due to an irregular potential landscape
produced by impurities, an asymmetric confinement
potential~\cite{jalabert92}, and/or electron-electron
interactions~\cite{mezamontes97}. The transport properties of quantum
dots are inherently related to their energy spectra and electronic
wavefunctions and thus the connection with RMT is readily
made~\cite{jalabert92,beenakker97}.

Indeed, experiments on the distribution of conductance peak heights of
quantum dots in the Coulomb blockade regime have shown good agreement
with the predictions of RMT~\cite{chang96,folk96}.  On the other hand,
% DA-M 1/14/99: has --> have
the distribution of the CB peak spacings have been found to deviate
from the expectations of RMT~\cite{sivan96,simmel97,patel98}.  The results
suggest that the peak spacings are not distributed according to the
famous Wigner surmise. Furthermore, there is no indication of spin
degeneracy which would result in a bimodal peak spacing distribution~\cite{patel98}.  
In Refs.~\cite{sivan96,simmel97} the
fluctuations of the peak spacings are considerably larger than
expected from RMT, whereas the experiments presented in
% DA-M 1/14/99: , added before & after however
Ref.~\cite{patel98} yield smaller peak spacing fluctuations, which,
however, are still larger than those predicted by RMT.

The deviations from the RMT predictions have been frequently
interpreted as fluctuations in the charging energy. As the charging
energy reflects the Coulomb interactions both between the electrons on
the dot as well as between the dot and its environment, the dependence
of the fluctuations on the interaction strength is of fundamental
interest. Numerical studies suggest that the fluctuations are
proportional to the charging energy rather than to the single particle
level spacing~\cite{sivan96,berkovits97,berkovits98}. This is also
found theoretically for the classical limit of a Coulomb glass
island~\cite{koulakov97}.  In RPA calculations the fluctuations have
been related to fluctuations of the eigenfunctions of the
dot~\cite{blanter97}. Another approach based on density functional
theory emphasizes the role of the Coulomb matrix elements of scarred
wavefunctions~\cite{stopa98}.  Recently, also a non-interacting
explanation for the Gaussian shape of the peak spacing distribution
has been given in terms of level dynamics due to shape deformation of
the quantum dot~\cite{vallejos98}.

Here we present an experimental study of the statistics of Coulomb
blockade peak positions of a quantum dot.  The dot is defined by
patterning the two-dimensional electron gas of a silicon metal oxide
semiconductor field effect transistor (MOSFET) structure using stacked
gates.  These experiments differ significantly in two major ways from
prior experiments on quantum dots defined in GaAs/AlGaAs
% DA-M 1/14/99: First --> first
heterostructures: first and foremost, due to the different electron
density and material properties of silicon, the ratio of the charging
energy, $E_C$, to the single particle energy level spacing, $\Delta
\epsilon$, is considerably larger; likewise the dimensionless
parameter $r_s$, which characterizes the strength of the Coulomb
interactions, is larger than in previous studies.  Secondly, the
number of electrons is varied by the application of a voltage to a top
gate instead of by squeezing the quantum dot with a plunger gate. We find
that the distribution of the peak spacings is unimodal and roughly
Gaussian. The magnitude of the fluctuations is 15 times larger than
that predicted by RMT.

Conduction through a small electron island coupled to leads via tunnel
barriers is normally suppressed if $k_B T \ll E_C$, where $E_C$ is the
charging energy of the island.  This effect is known as the Coulomb
blockade~\cite{cbreview}. The blockade is lifted when the condition
% DA-M 1/14/99: 'and' between mu_d and mu_{dot} (& changes in math mode)
% DA-M 1/14/99: 'the' before source & removed before dot
$\mu_d < \mu_{dot} < \mu_s$ is satisfied, where $\mu_s$, $\mu_d$, and
$\mu_{dot}$ are the chemical potentials of the source, drain and dot,
respectively. In the linear response regime, where these experiments
have been performed, $| \mu_d-\mu_s | \ll \Delta \epsilon ,k_BT$.  The
chemical potential of the dot is defined as
$\mu_{dot}(N+1)=E(N+1)-E(N)$ where $E(N)$ is the total energy of the
% DA-M 1/14/99: with $N$ --> by $N$
% DA-M 1/14/99: In this case --> In the case where the blockade is lifted...
dot occupied by $N$ electrons.  In the case where the blockade is lifted an electron can tunnel
from the source onto the dot, changing the dot's occupation from $N$
to $N+1$, and sequentially tunnel off the dot to the drain leaving the
dot in its original state. The resulting fluctuation of the electron
number on the dot leads to a finite conductance.  Experimentally this
can be achieved by appropriately tuning $\mu_{dot}$ with an external
gate. Sweeping the gate voltage, $V_g$, results in the well known
conductance oscillations indicating successive filling of the dot with
single electrons. The difference of $\mu_{dot}$ between two adjacent
conductance maxima is thus given by $\Delta \mu_N=E(N+1) - 2 E(N) +
E(N-1)$ which can be viewed as the discrete second derivative of the
quantum dot energy with respect to particle number, i.e.  the inverse
compressibility $\partial \mu/\partial N$~\cite{berkovits97}.

In the constant interaction (CI) model~\cite{cbreview} the energy of
the dot is approximated as $E(N)=(Ne)^2/2C_\Sigma + \sum_{i=1}^N
\epsilon_i$, where the electrostatic interactions are treated using a
simple capacitive charging model with a total dot capacitance
$C_\Sigma$, and the quantum mechanical terms are taken into account as
single particle energies $\epsilon_i$.  In this model the difference
of the chemical potentials for successive occupation numbers, the so
called addition energy, is $\Delta \mu_N=E_C+\Delta \epsilon_N$ with
the charging energy $E_C=e^2/C_{\Sigma}$, and the level spacing
$\Delta \epsilon_N=\epsilon_{N+1}-\epsilon_N$. This is mapped to gate
voltages via $e \frac{C_g}{C_{\Sigma}} \Delta V_g=E_C + \Delta
\epsilon_N$ where $C_g$ is the capacitance of the dot to the gate and
$\Delta V_g$ the difference between the gate voltages at which
adjacent conductance maxima occur.

% DA-M 1/14/99: 'has' deleted in 'has motivated...'
This final expression motivated the original investigations of the
peak spacings in the light of random matrix theory. RMT shows that the
normalized spacings $S$ ($\left<S\right>=1$) between adjacent
eigenvalues of a generic time-reversal invariant Hamiltonian are
distributed according to the Wigner surmise
\begin{equation}
P_W(S)=\frac{\pi}{2}  S e^{-\frac{\pi}{4} S^2}.
\label{wigner}
\end{equation}
The fluctuations of these spacings are $(\left< S^2 \right> - \left< S
\right>^2)^{1/2} \approx 0.52 \left<S\right>$.  However, experiments
have shown that the combined CI-RMT model is not capable of describing
the observed peak spacing distribution
correctly~\cite{sivan96,simmel97,patel98}.

The Coulomb blockade measurements on which the following analysis is
based have been performed on a quantum dot defined in a silicon MOSFET
structure.  We have utilized a stacked gate structure to pattern the
electron gas as shown in Fig.~\ref{structure}.  First, a gate oxide is
grown on a p-type silicon substrate (lower oxide), and then a lower
metal gate is deposited and patterned using electron beam lithography
and lift-off techniques.  Above the lower gate a second layer of
silicon dioxide is deposited (upper oxide), and finally an upper gate
is formed; the upper oxide layer serves to insulate the lower gate
from the upper gate.  Application of positive voltages to the upper
gate leads to the formation of a two-dimensional electron gas (2DEG)
at the Si/SiO$_2$ interface; $n^+$ implanted regions serve as Ohmic
contacts to the 2DEG.  Further details about this device may be found
elsewhere~\cite{abusch-magder96,abusch-magder97}.  The lower gates
locally screen the electric field created by the upper gate, and a
% DA-M 1/14/99: deleted 'an' in an appropriate...
quantum dot is formed by applying appropriate negative voltages to the
% DA-M 1/14/99: the sentence The size... is rewritten was The lithographic...
lower gates.  The size of the dot is estimated from the capacitance to
be $A \approx 200\,\rm{nm} \times 200$\,nm, which agrees well with the
lithographic dimensions of $250\,\rm{nm} \times 270$\,nm when
electrostatic depletion at the edge is considered.  The electron
density can be varied by changing the upper gate voltage, whereas the
lower gate voltage controls the tunnel barriers and the electrostatic
confinement potential of the quantum dot. This technique allows the
definition of very small structures which therefore have low
capacitances and high charging energies.  For the quantum dot
discussed here these values are $C_\Sigma \approx 85$ aF and $E_C
\approx 1.9$ meV as obtained from temperature dependence measurements
of the conductance resonances~\cite{abusch-magder97}.

In contrast to previous experiments on quantum dots in
GaAs/AlGaAs heterostructures the electron density is considerably
higher, $n_s \approx 2.5\cdot 10^{16}\,\rm{m}^{-2}$.  The mobility of the
two-dimensional electron gas is $\mu=0.56\,\rm{m}^2/$Vs, and the mean free
path $l \approx 100\,$nm is comparable to the system size.  The single
particle energy level spacing can be obtained from the estimated dot
% DA-M 1/14/99: put ,s around $A$
area, $A$, via Weyl's formula~\cite{baltes76} as $\Delta \epsilon =
\frac{2\pi \hbar^2}{gm^* A}= 15 \,\mu$eV, where $g$ is the degeneracy
of electronic states in the two-dimensional electron gas, and $m^*$ is
the effective mass of the electrons.  In a 2DEG in a silicon MOS system
$m^* = 0.2 m_e$, and at $B=0$ both the spin and valley degeneracies
must be considered and therefore $g=4$.  While these quantum dots are
smaller than many of the GaAs/AlGaAs quantum dots
studied~\cite{sivan96,simmel97,patel98}, $\Delta \epsilon$ is of the
same order due to the larger effective mass, and to the valley degeneracy.

The strength of the electronic interactions characterized by the
% DA-M 1/14/99: deleted Sivan ref, this is generally known
dimensionless parameter $r_s=g/(2 \sqrt{\pi n_s} a_B^*) =
2.1$ considerably exceeds the values obtained in recent
experiments (where $r_s \approx 1$)~\cite{sivan96,simmel97,patel98};
here $a_B^*$ is the effective Bohr radius.  Similarly, the ratio of
charging energy to single particle energy level spacing $E_C/\Delta
% DA-M 1/14/99: added note: ', another measure... interactions,'
\epsilon \approx 125$, another measure of the relative importance of
electron-electron interactions, is larger than in the experiments performed on
GaAs/AlGaAs quantum dots.

The measurements were performed in a $^3$He refrigerator at a
temperature of $T=320$mK using standard lock-in techniques at low
frequencies and bias. The conductance oscillations were measured as a
function of the upper gate voltage. Consequently, the electron density
was varied without drastically changing other system parameters such
as charging energy, single particle energy, and dot shape.  This also
contrasts with former experiments on the statistics of conductance
oscillations where the shape of the quantum dot was distorted by
plunger gates~\cite{folk96,patel98}.

The following analysis is based on a series of more than $100$
conductance peaks occurring in the upper gate voltage range from
$12.1$ V to $13.5$ V (see Fig.~\ref{fluctuation} inset).  In this
range the electron density changes from $2.4 \times 10^{16}\,\rm{m}^{-2}$
to $2.6 \times 10^{16}\,\rm{m}^{-2}$.  The quantum point contacts
connecting the quantum dot to the leads are tuned into the tunneling
regime by applying voltages of $-4.5$\,V and $-8.0$\,V to the left and
right pair of lower gates, respectively.  The position
% DA-M 1/14/99: added: 'by fitting the peak by a...'
of each peak is obtained by fitting the peak by a thermally broadened
line shape $\propto \cosh^{-2}(e \alpha (V-V_0)/2 k_B
T)$~\cite{cbreview} where $\alpha$ and $V_0$ serve as the fit
parameters.  The gate voltage spacings $\Delta V_g$ are
calculated from the peak positions.

The mean value $\left <\Delta V_g \right >$ of the spacings as a
function of $V_g$ is obtained from a linear fit as $\Delta V_g (V_g)
\approx \left[ 12.2 - \frac{3.5 \cdot 10^{-2}}{\rm{V}} \times
(V_g-12.5\,\rm{V}) \right] \,$mV.  The smallness of the slope of this
fit shows that the influence of the upper gate on the capacitance and
% DA-M 1/14/99: therefore --> therefore on the
therefore on the size of the dot is rather weak.  Accordingly, the shape
deformation which has been postulated to explain the distribution of
% DA-M 1/15/99:  'of the $\Delta V_g$' --> 'of $\Delta V_g$'
$\Delta V_g$~\cite{vallejos98} plays no significant role in this
experiment.  The normalized peak spacings
\begin{equation}
\delta=\frac{\Delta V_g - \left< \Delta V_g \right>}{\left< \Delta V_g \right>}
\end{equation}
are displayed in Fig.~\ref{fluctuation}.  The fluctuations of $\delta$
are computed to be $\left<\delta^2\right>^{1/2} \approx 0.06$.  The
fluctuations in the addition energy are, therefore, roughly $115
\mu$eV, which is 7.5 times the mean level spacing $\Delta \epsilon$,
and thus 15 times larger than the fluctuations expected from RMT. This
supports the view that the fluctuations of the addition energy scale
with the Coulomb energy rather than with the kinetic energy.  However,
the proportionality factor 0.06 is smaller than that suggested by
numerical calculations (0.1 - 0.2)~\cite{berkovits98}.  It should be
noted that in these experiments $\Delta \epsilon \approx k_B T$.  We
expect that the effect of thermal broadening would be to reduce the
fluctuations in peak spacing.  A simple model~\cite{patel98} predicts
% DA-M 1/14/99: 'predicted by RMT' --> 'expected within RMT'
% DA-M 1/14/99: 'into the predictions of RMT then' --> 'into RMT then'
that the fluctuations expected within RMT would be reduced by a factor of
2 -- 3.  If we incorporate this correction into RMT
then the fluctuations we find in our experiment are 30 -- 45 times
larger than those predicted by RMT.

% DA-M 1/15/99: made sure there was a paragraph break in preprint form
The distribution of the peak spacings normalized to an area of unity
is shown in Fig.~\ref{histogram}.  The distribution is unimodal and
roughly has the shape of a Gaussian. In the inset of
Fig.~\ref{histogram} the experimental distribution is depicted
together with the Wigner surmise (Eq.~\ref{wigner}); for comparison to
$\delta$ we have rescaled the predictions of RMT taking into account
the experimental values of $E_C$ and $\Delta \epsilon$.  As in
previous experiments there is no evidence of a bimodal addition
spectrum as is predicted by the CI-RMT model. This is in agreement
with the theoretical prediction that the influence of spin degeneracy
on the addition spectrum is washed out for stronger electron-electron
interactions ($r_s > 1$)~\cite{berkovits98,blanter97}.

In conclusion, we have investigated the Coulomb blockade peak spacing
distribution of a quantum dot fabricated in the 2DEG of a silicon
MOSFET structure. In accordance with experiments on GaAs/AlGaAs
quantum dots the distribution differs from the Wigner surmise and is
roughly Gaussian. The fluctuations are approximately $0.06 \times
E_C$.  Due to the large ratio of charging energy $E_C$ to single
particle energy $\Delta \epsilon$ this strongly suggests that the
fluctuations scale with $E_C$ and not with $\Delta \epsilon$. This
clearly contradicts the predictions of RMT and indicates that the
fluctuations are dominated by electron-electron interactions in this
system.

* new address: 
Bell Laboratories, Lucent Technologies, 600 Mountain Ave., Murray Hill NJ 07974 

\acknowledgments
The authors would like to thank the staff of the
Microelectronics Lab at Lincoln Laboratory for their help with device
fabrication, and U. Sivan, M. Stopa, H. Baranger for fruitful
discussions.  This work was supported by the Alexander-von-Humboldt
Foundation, the DFG Sonderforschungsbereich 348, and by the Army
Research Office under contract~DAAH04-94-0199 and
contract~DAAH04-95-0038.  Work at Lincoln Laboratory was sponsored
by the U.S. Air Force.  DA-M gratefully acknowledges support from the
Alexander-von-Humboldt Foundation, and the Bell Labs Foundation.

\begin{figure}
%\begin{center}
%\epsfig{file=fig1.EPS,width=6.5cm}
%\end{center}
\caption{
Schematic representation of the device design. A
cross section of the device is shown in (a).  Two oxide and two gate
layers are formed on top of p-type silicon substrate.  The lower oxide
layer has a thickness of roughly 20 nm, while the upper oxide is
approximately 80\,nm thick. The voltage on the upper gate is used to
vary the electron density in the 2DEG induced at the interface of
the lower oxide and the silicon. A top view of the device is shown in
(b).  The pattern in the lower gates defines a quantum dot in the
induced electrons; note that the upper gate covers all of the area
show in (b) and overlaps the source and drain.  The lithographic
dimensions of the quantum dot are 250 nm $\times$ 270 nm.}
\label{structure}
\end{figure}

\begin{figure}
%\begin{center}
%\epsfig{file=fig2.EPS,width=9cm}
%\end{center}
\caption{The normalized peak spacings $\delta$ obtained from an upper
gate voltage sweep.  The fluctuations around the mean value $0$ are
much larger than expected from the CI-RMT model.  The inset shows
Coulomb blockade conductance oscillations as a function of the upper
gate voltage.}
\label{fluctuation}
\end{figure}

\begin{figure}[b]
%\begin{center}
%\epsfig{file=fig3.EPS,width=9cm}
%\end{center}
\caption{Histogram showing the distribution of the normalized peak
spacings from Fig.\ 2.  The area of the histogram is normalized to
unity. A Gaussian fit with standard deviation of $\sigma= 0.06$ is
also displayed.  The inset shows the same histogram alongside the
Wigner surmise, the distribution predicted by RMT, for $\Delta
\epsilon = 15\,\mu$eV.  The experimental distribution is much broader
than expected from RMT.}
\label{histogram}
\end{figure}

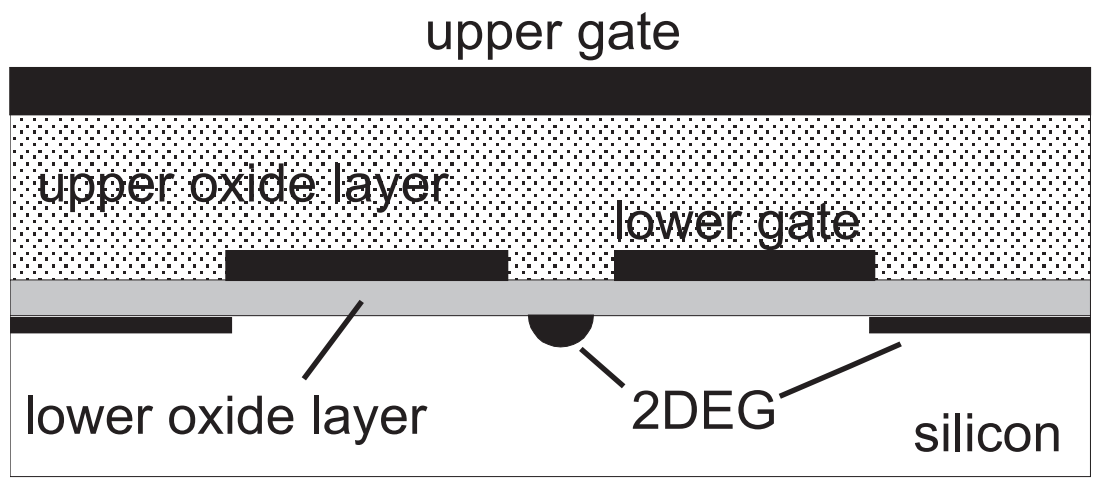
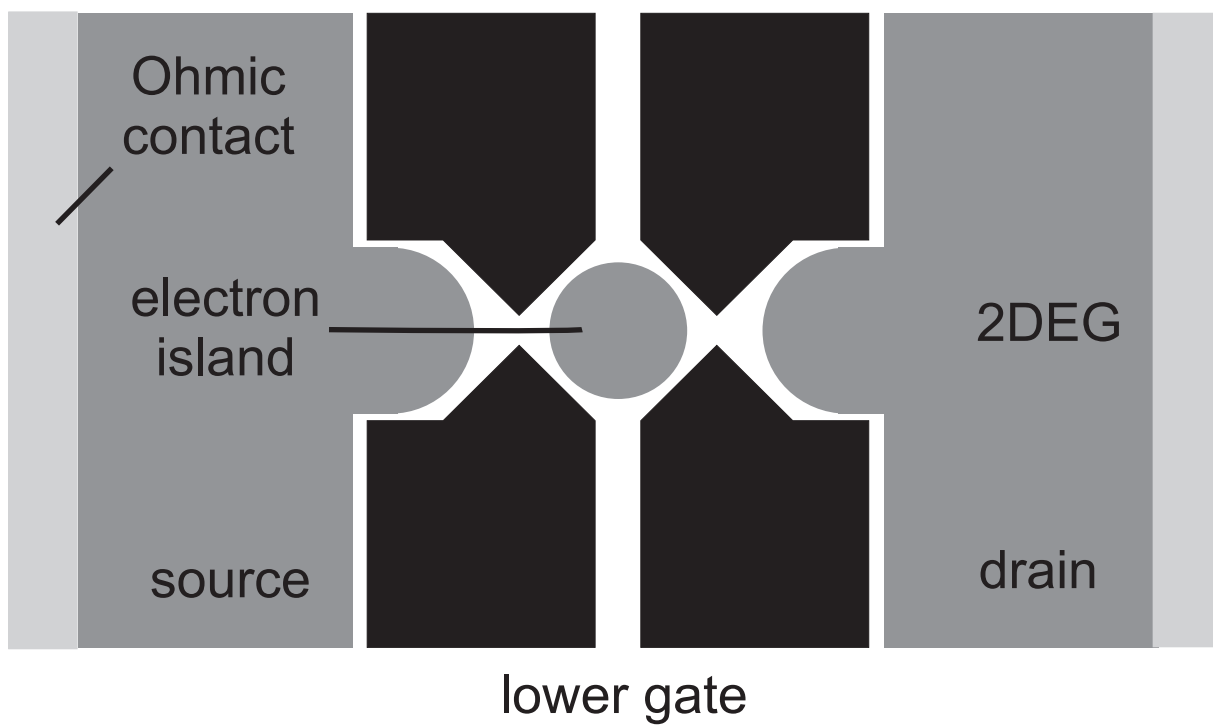
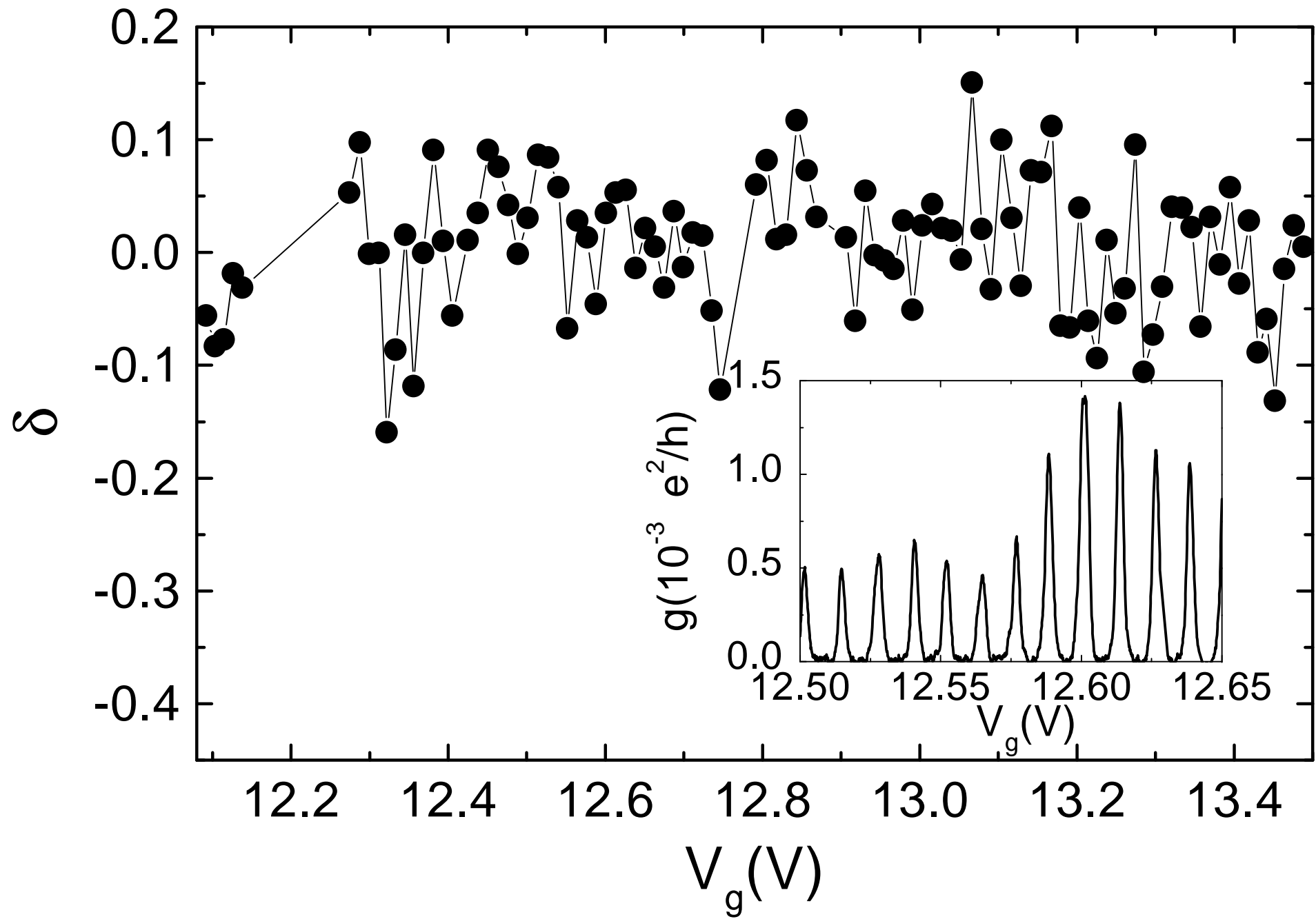
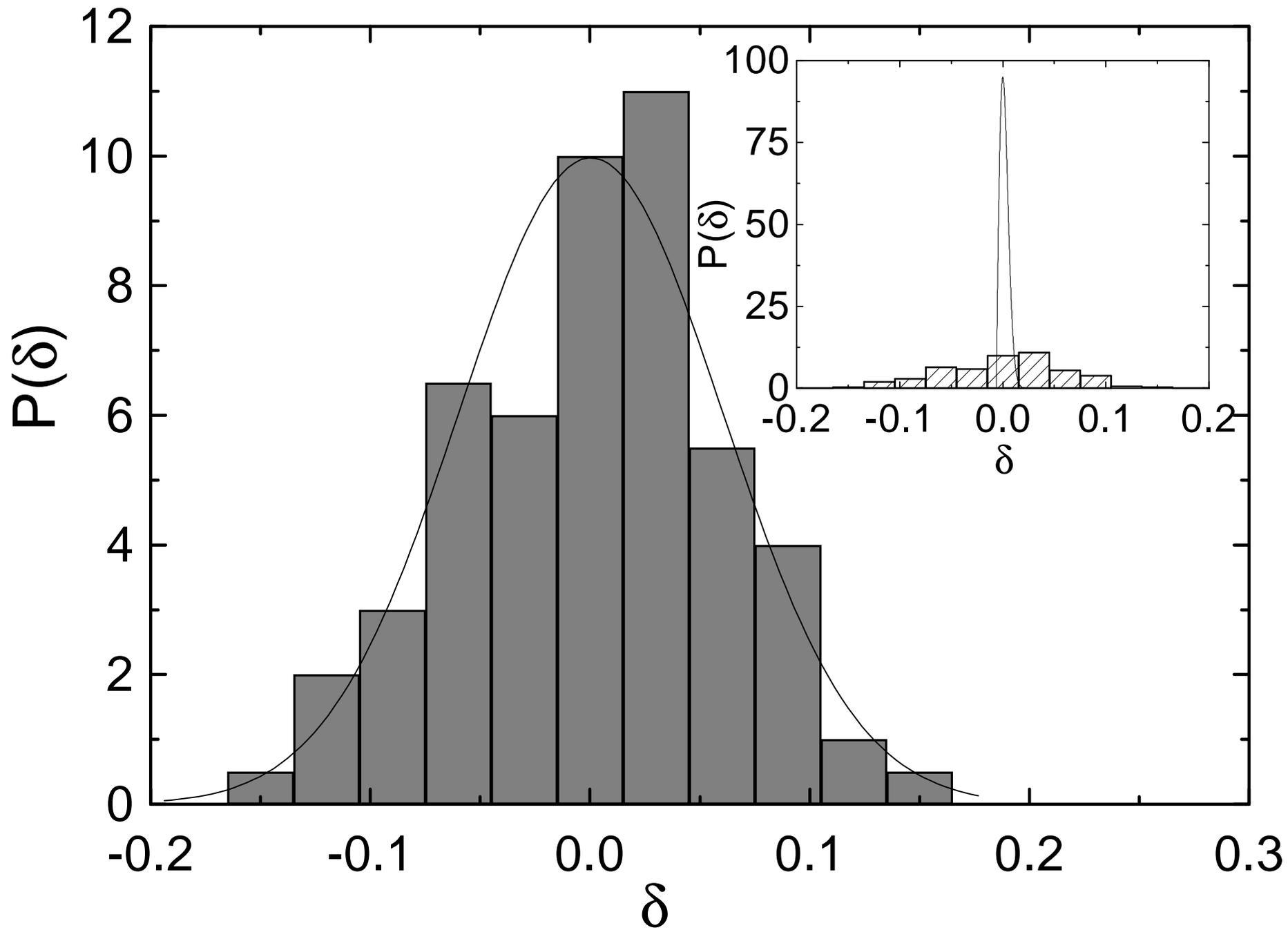
\end{document}